\documentclass[preprint,showpacs,preprintnumbers,amsmath,amssymb,floatfix]{revtex4}
\usepackage{graphicx}
\usepackage{dcolumn}
\usepackage{bm}
\usepackage{longtable}
\begin{document}

\title{ Bulk and nano GaN: Role of Ga $d$ states}
\author{Roby Cherian, Priya Mahadevan}
\affiliation{ S.N. Bose National Centre for Basic Sciences, JD-Block, Sector III,Salt Lake, Kolkata-700098, India }
\date{\today}

\begin{abstract}

We have studied the role of Ga 3$d$ states in determining the properties of
bulk as well as nanoparticles of GaN using PAW potentials. 
A significant
contribution of the Ga $d$ states in the valence band is found to 
arise from interacion of Ga 4$d$ states with the dominantly 
N $p$ states making up the valence band.
The errors arising from not treating the Ga 3$d$ states as a part of the
valence are found to be similar, $\sim$ 1$\%$, for bulk as well as for nanoclusters of GaN.

\end{abstract}

\maketitle

{\bf I. Introduction}

The semiconductor industry has used silicon as its basic building
block since the beginning. Recently the focus has shifted to other
materials with the aim of harnessing their multifunctionality to
construct new generation devices. An important class of materials
that have received considerable attention in this context are the
nitrides. The bulk nitrides formed with group III elements show
interesting physical properties such as large piezoelectric response
\cite{piezo}. However, the dilute nitrides where one observes
dramatic effects by the introduction of a small amount of nitrogen
to replace the anion sites have received a lot of attention in
recent times. Alloys of GaInNAs have been recognized as important
materials for the development of long wavelength solid-state lasers
emitting within the fibre-optic communication wavelength window (1.3
to 1.55 $\mu$m) \cite{gainas}. There are also theoretical
predictions that exist which suggest that these materials could also
be used to increase the efficiency of multijunctional solar-cells
\cite{solar-cell}. In the case of GaNP alloys, the crossover from an
indirect to a direct bandgap induced by N incorporation
\cite{direct} promises high radiative efficiency, whereas a
N-induced reduction in the lattice constant offers a possibility of
lattice matching between optically efficient III-V compounds and Si
wafers, desirable for the integration of the two technologies
\cite{si-integrate}. GaInNAs semiconductor quantum dots with dilute
amount of nitrogen substitutional impurities are promising
candidates for the active region in the next generation of
optoelectronic devices \cite{qdots}. Transition metal doped GaN has
been found to exhibit ferromagnetism at room temperature \cite{reed}
which could make these materials useful in the emerging area of
spintronics.

With recent advances in computational power, theory and specifically
ab-initio density functional theory has played an important role in
the design of materials with tailor-made properties \cite{dft}. Calculations
for the systems of interest in the context of the nitrides - dilute nitrides 
as well as quantum dots, are usually performed for periodic systems 
considering large and representative 
supercells. These are computationally demanding within
an ab-initio approach.  It is
therefore useful to have accurate and reasonable approximations
which decrease the computational cost. In this context it was shown
that enormous saving in terms of computational time may be made if
one used ultrasoft pseudopotentials developed by Vanderbilt
\cite{vanderbilt}. Further improvements were made by Bl$\ddot{o}$chl
\cite{blochl} who combined the ideas of soft pseudopotentials and
all electron methods like LAPW (linearised augmented plane wave)
within an elegant framework called the projected augmented wave
(PAW) method. In this work we have examined the bulk electronic
structure of GaN using PAW potentials. The results have been compared
with those obtained using ultrasoft pseudopotentials. The calculated 
equilibrium lattice constants are within 0.3 $\%$ of each other. 

The role of the Ga 3$d$ states in determining the physical properties
of GaN has received considerable attention over the past two decades. Unlike in the case of other
Ga-V semiconductors one finds that in GaN the Ga 3 $d$ core states are not chemically inert. 
One finds a strong overlap in energy between the
semi-core Ga 3$d$ states as well as the N 2$s$ states. Fiorentini et al. \cite{fiorentini} pointed out that 
ignoring this interaction would have consequences on both the cohesive energy as well as the
equilibrium lattice constant deduced theoretically. According to the variational principle, neglect 
of any relaxation of the semi core levels would increase the total energy, an effect which would disappear
in the limit of well separated atoms in the solid. Further, the overlap of the core states with the 
valence states on the same atom results in a nonlinear exchange interaction. Finally the interaction of core states
with core states on the neighboring atom results in the neglect of closed shell repulsion. This
has the effect of an increase in the kinetic energy when the cores on neighboring atoms are made orthogonal. If this
contribution is neglected, the cohesive energy comes out too large and the lattice constant too small. 
The net effect which they found was that the lattice constant when Ga 3$d$ states were neglected was almost
4 $\%$ smaller than that obtained with the Ga 3$d$ states included within LDA. An additional effect of the
neglect of the Ga 3$d$ states is observed in the modification of the band gap.
The Ga 3$d$ states that split 
into states with $t_2$ and $e$ symmetry in the tetrahedral crystal field of the N neighbors, interact with the valence 
band derived levels with the same symmetry. This interaction pushes the valence band maximum to 
higher energies and therefore decreases the band gap of GaN in comparison to treatments in which
Ga 3$d$ states were a part of the core. 
Recent studies  by Bouhafs et al. \cite{bouhafs} on the GaN semiconductor, at 
a fixed lattice constant, also confirm that the bandgap decreases 
in a treatment in which the Ga 3$d$ states were included in the valence.

PAW potentials give us the freedom to toggle between using the Ga 3$d$ in the
valence and in the core and allows us to simultaneously examine the modifications in the electronic
properties and more importantly the structural and cohesive properties. 
The implementation of PAW that we use allows for core-valence interaction 
within a frozen core approximation.
We first review the existing understanding for bulk GaN using PAW potentials. The equilibrium lattice
constant computed by us within pseudopotential calculations with and without Ga 3$d$ in the valence differ by
less than 0.3 $\%$ using ultrasoft pseudopotentials. The deviations between the two approaches is similar when
we use PAW potentials. All earlier studies have found that the lattice constant without Ga 3$d$ in the valence 
is underestimated within the pseudopotential approach, and our results are consistent with this observation.
The PAW approach gives us a different trend, however, and we attribute that to the treatment of core-valence exchange 
interaction. Changing the anion to P and then As, we find an overestimation of the lattice constant when Ga 
3$d$ states are not included as a part of the valence. The difference between the theoretical lattice constants, 
however, decreases as we go from GaN to GaAs. A considerable portion of the 
literature has commented on the Ga 3$d$ admixture in the valence band of GaN.
To explicitly examine this, we have plotted the Ga $d$ partial density of states 
for both cases - with and without Ga 3$d$ states in the valence. The Ga $d$ contribution 
in the valence band arising from semi-core valence interaction accounts for  $51\%$ of the total 
$d$ admixture. This ratio decreases as we move to GaP and GaAs.

Having studied the bulk limit of GaN, we examined small representative clusters of GaN. Quantum 
confinement effects modify the energy of the valence band maximum and conduction band bottom of 
the semiconductor nanoparticles, and should decrease the
separation between the semi core Ga 3$d$ states and the valence band maximum. This results in an 
increased interaction strength and therefore an enhanced 3$d$ contribution in the valence band.
Comparing the equilibrium lattice constant with and without Ga 3$d$, we find a difference of $\sim$ $1\%$
for clusters with an average diameter of $\sim$ 10 $\AA$.

{\bf II. Methodology}

The electronic structure of bulk zinc-blende GaN, GaP and GaAs was
calculated using a plane wave pseudopotential 
implementation of density functional theory within VASP \cite{vasp}. 
Ultrasoft pseudopotentials as well as PAW potentials \cite{paw} have been used. 
Calculations have been performed with and without Ga 3$d$ states included
in the valence band. GGA-PW91 approximation  \cite{ggapw91} has been used for the exchange.
A dense k-points grid  of 8x8x8 within Monkhorst Pack scheme has been used. 
The energy cutoffs used for the kinetic energy of the plane waves used in the
basis was 500 eV for GaN, 337.5 eV for GaP and 260.9 eV for GaAs. The calculations
for GaP and GaAs which did not include the Ga 3$d$ states as a part of the valence 
band had a cutoff of 353.4 eV. The convergence with respect to k-points was tested
by increasing the mesh density from 8x8x8 to 10x10x10. The total energies changed by 
0.02 meV. The equilibrium lattice constant has been determined by fitting the energy variation
with volume to the Murnaghan equation of state. An analysis of the electronic structure
was performed using the density of states calculated using the tetrahedron method. The
wavefunctions were projected onto atom-centered spherical harmonics integrated over 
spheres of radii 1.2 $\AA$ for Ga, P and As in GaP, GaAs and GaN and 0.95 $\AA$ for N 
in GaN for the evaluation of the density of states.

We also examined the electronic structure of GaN nanocrystals 
in the cluster limit
by considering representative clusters. 
We construct nanocrystals by cutting a spherical fragment of a
bulk crystal, which has an underlying geometry of the zincblende structure.
Now to define a spherical nanocrystal in this way we need to specify
the center and the radius. In our studies the nanocrystal is centered
on the Ga atom, and then the nanocrystals are generated by considering a 
spherical cut off radius. These will have a T$_d$ point group symmetry.
The smallest cluster considered had 4 atoms around the central Ga atom, 
and since it had just one layer around the central atom for 
simplicity we denote this cluster as n=1 (where n stands for the number
of layers around the central atom). The next size cluster which was considered 
in our study had 3 layers around the central 
atom (n=3), having in total 13 Ga and 16 N atoms (Fig.1).

Calculating the equilibrium lattice constant of the cluster is a global
optimization problem. Instead of allowing all degrees of freedom to be 
optimized simultaneously, we carried the optimization in parts.
The cluster was expanded and contracted
keeping the geometry fixed, {\it i.e. allowing for a volume dilation/contraction} 
about the bulk like fragments. At each size the convex hull 
formed by the surface atoms was constructed and it was used 
to compute the volume. The equilibrium lattice 
constant was then calculated by fitting the energy variation with 
volume to the Murnaghan equation of state \cite{murnaghan}.
The clusters were then hydrogenated using pseudo hydrogens and the
atom positions were optimised to reach minimum force positions. An 
average bondlength was determined by averaging over all the 
nearest-neighbor bondlengths. This was then used to determine an average
equilibrium lattice constant.
Again as done in the case of the bulk, the equilibrium lattice constant with and
without Ga 3$d$ states in the valence were determined. 
A similar analysis was performed for a cluster with three shells (n=3). 
Features of the
electronic structure are examined by calculating the density of states
broadening each eigenvalue with a gaussian of full width at half maximum
of 0.1~eV.

{\bf III. Results and Discussion}

As discussed earlier , the near resonance of the Ga 3$d$ states with the N 2$s$ states
results in a strong deviations in calculated structural properties in treatments where
Ga 3$d$ states are not included as a part of the valence band. These considerations
prompted us to carry out calculations using PAW potentials, allowing us to toggle between
using Ga $d$ in the valence, and merely as a part of the core. 
The results are given in Table I. For the
comparison the results using ultrasoft potentials were also calculated (Table I).
The error in the calculated lattice constant with and without  $d$ states in the
valence were $\sim$ 0.03-0.04 $\AA$ (around 1$\%$). A smaller error in the calculated
lattice constant is also found when one used ultrasoft potentials with and without
Ga $d$ in the valence. These results suggest that possibly the large deviations in the equilibrium 
lattice constant found earlier are specific to the choice of the method. The trends in the lattice constant 
with and without $d$ are in opposite directions when we used ultrasoft potentials
and when we use PAW potentials. As the
treatment of the core electrons are meaningful in the PAW calculations,
we examined these calculations in greater detail.
The equilibrium lattice constant is predicted to be smaller when Ga $d$ states are
included in the valence. This is a suprising result at first as Ga $d$ states interact
primarily with the filled N $s$ and N $p$ states in the valence band. Hence, naively
one does not expect there to be any energy gain as a result of the interaction.
However the valence and conduction band electrons feel the presence of the
Ga 3$d$ electrons in the semi-core. Our recent analysis \cite{rc-pm-cp} has
shown the manner in which the Ga $d$ states interact with
valence band states. By artificially moving the Ga $d$ states to deeper energies
using a $U$  on the 3$d$ states within the framework of LDA+U, we simulated the
situations of having / not having chemically active Ga 3$d$
states. Gradually moving the Ga 3$d$ states to deeper energies we find a redistribution
of charge on Ga related levels. This in turn leads to a modification of the
interaction between the anion $p$ states and cation states. The altered interaction
strengths can therefore explain why there should be any modification of the total
energy and therefore the lattice constant of these systems with and without
the inclusion of Ga 3$d$ states in the valence.

Moving down the Group V series of the periodic table to heavier anions instead of Nitrogen, we find a
similar trend. The theoretical lattice constant (Table II) calculated 
within the PAW method in the absence of 3$d$ in
the valence for Ga are consistently larger than when the 3$d$ states are treated as
a part of the valence. With increasing atomic number on the anion, the Ga 3$d$ states are
pushed deeper into the valence band, and hence their interaction with the anion $p$ states
making up the valence band are weaker. Hence the deviation in the equilibrium lattice constant with
the two choice of basis becomes smaller as we go from GaP to GaAs. While the deviations in the 
theoretical lattice constant are small, the errors in the theoretical bulk modulus are 
significant in the case of GaN, while they are small in the case of GaP and GaAs.

The significant interaction between the Ga 3$d$ states with the N $p$ states comprising the valence
band is usually measured by plotting the Ga $d$ admixture in the valence band. Our choice of basis,
however, allows us to distinguish the 3$d$ admixture from the 4$d$ admixture, which one believes
is not strongly affected by changing the basis and is largely additive.
The total as well as the s,p,d contribution to the Ga and N partial density of
states have been plotted (Fig.2) for GaN with the 3$d$ states on the Ga treated as a
part of the core. The zero of the energy axis has been set to be the valence band
maximum. The N $s$ states contribute at around -11.5 eV while the N $p$ states
contribute between 0-6 eV. The band gap is calculated to be 1.47 eV within the
present calculation. Ga $s$ and $p$ states are strongly intermixed in the
conduction band. As is evident from the middle panel, there is a small
admixture of the Ga 4$d$ states within the states comprising the valence band
(especially 0-3 eV below the valence band maximum).

A similar plot (Fig.3) has been made from the calculations which include Ga 3$d$
states in the valence. The gross features of the electronic structure remain
unchanged. The Ga 3$d$ states are found to lie at higher energies in these calculations
than the N $s$ states. Significant interaction is found to exist between the semi core
N $s$ and Ga $d$ states because of their close proximity in energy. The Ga $d$ states
in the semi core also interact with the N $p$ states. The bandgap in the current
calculation is found to be 1.56 eV, therefore increased by $\sim$ 90 meV from
the value obtained when the Ga 3$d$ states were a part of the core. It should
be noted that the density of states have been plotted at the theoretical 
equilibrium lattice constants given in Table I. Had we fixed the lattice constant in the two
calculations, we would have seen a reduction in the band gap when the Ga 3$d$ states were
included in the basis as observed earlier \cite{bouhafs}.
Here we have the additional effect of a decreased lattice constant
and so we find a larger band gap. 

We have also examined the change in Ga $d$ contribution in the valence and conduction band with the 
two choice of basis. This is plotted in Fig. 4. Assuming that the Ga 4$d$ admixture in the
valence band is unchanged when Ga 3$d$ states are included in the basis, the results are quite surprising.
We find that the Ga 3$d$ admixture in the valence band accounts for around 51$\%$ of the total Ga $d$
component in the valence band. This is contrary to the belief that the Ga $d$ contribution in the valence band
is a measure of the semi-core valence interaction. Similar results are plotted for GaP and GaAs
in Figs. 5 and 6 at their theoretical equilibrium lattice constants (Table II). The $d$ admixture gradually decreases
as we go from GaN to GaP and finally to GaAs, and is mainly from interaction of the 
anion $p$ states with the Ga 4$d$ states in the conduction band. The Ga 3$d$ admixture in the 
valence band accounts for around 42$\%$ and 23$\%$ of the total Ga $d$ component in the valence band for GaP 
and GaAs respectively.

As GaN showed significant interaction between the Ga 3$d$ states with the N $p$ sates,
we examined the modifications in the interactions and consequent implications when
one went down to the nanoregime. As is well known, quantum confinement effects modify
the position of the levels which move from their positions in the bulk solid to
deeper energies at a rate which is inversely proportional to the effective mass of the
level. Since the $d$ states would move more slowly than the states comprising the valence
band maximum, with decreased cluster size, one expects the Ga $d$ - N $p$ seperation to
decrease, and hence interaction to increase. Indeed this is found to be the case,
and one measures the enhancement in the $p$-$d$ strength by the relative error that
one finds in computed quatities such as the lattice constant. In Table III we
provide the optimised lattice constants for the two representative clusters. These are
found to be smaller than that for the bulk GaN. As the size of the cluster 
decreases we find the the relative position of the Ga 3$d$ from the valence band 
maximum to decrease, for the smallest cluster (n=1) the seperation is reduced by 2 eV 
and for the n=3 case it is reduced by 0.6 eV, with respect to the bulk separation value,
resulting in the increased $p$-$d$ interaction which modifies the lattice constant.
With the two choices of basis we also examined the changes in the
Ga $d$ and N $p$ contribution in the valence and
conduction band. Around the conduction band region the changes resulting from 
the choice of the two basis were small. For the two nanocluster cases (n=1 and n=3) 
the density of states around the valence band region are shown in Fig. 7. The 
zero of the energy corresponds to the valence band maximum.
Here the Ga 3$d$ admixture in the valence band accounts for around 53$\%$
for the n=1 case and 51$\%$  for the n=3 case of the total Ga $d$ component
in the valence band, which is almost the same as what 
we had observed for the bulk.
Further the presence and absence of the semi-core Ga 3$d$ states modifies the lattice
constant in the same direction as the bulk calculations. The deviations are found to 
of the same order as that observed for the bulk. At the theoretical calculated 
equilibrium lattice constant of these nanoclusters we found bandgap of 5.45 
and 5.46 eV within our calculations and larger cluster had a bandgap of 4.79 and 4.76 eV, for the cases 
with and without the inclusion of Ga 3$d$ states in the basis seperatively.

{\bf IV. Conclusion}

Hence we have studied the modification of the equilibrium properties for GaN,
with and without treating the Ga 3$d$ in the valence, in both the bulk as well
as in the cluster limit. The effects of the lattice constant modification are found to 
be small and of the order of 1$\%$ at both limits. Hence we conclude that a
treatment using PAW potentials where Ga 3$d$ states are treated 
as a part of the core 
is adequate to describe the properties of the GaN.

\renewcommand

\newpage
\begin{figure}
\includegraphics[width=5.5in,angle=000]{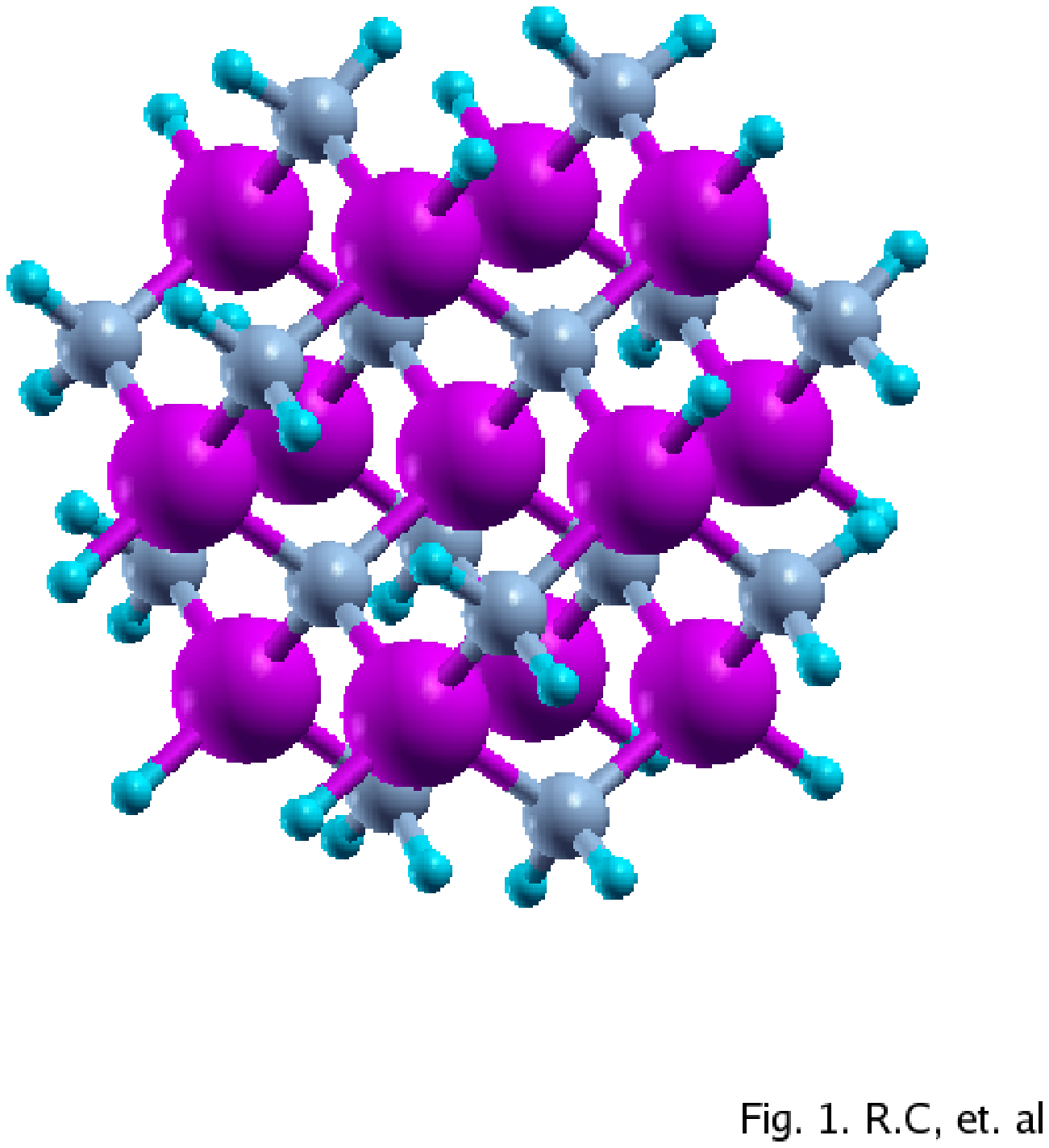}
\caption{(Color online). A ball and stick model for spherical nanocrystals 
(n=3) of GaN having
an underlying  zinc blende geometry.  The dark violet
balls correspond to Ga atoms, the light grey balls correspond
to N atoms and the outer smaller blue balls  denote 
the pseudo-hydrogen atoms.}
\end{figure}

\begin{figure}
\includegraphics[width=5.5in,angle=270]{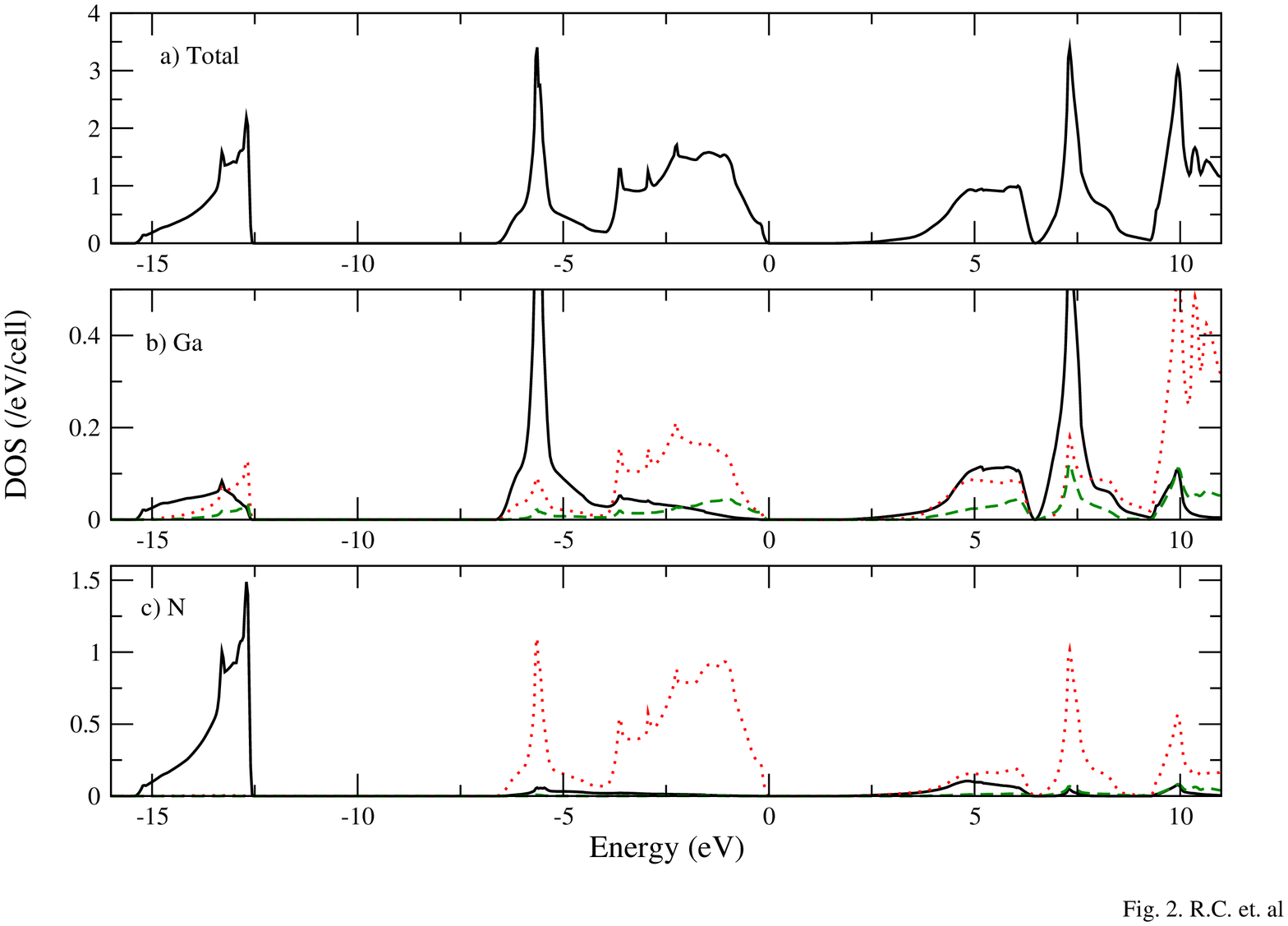}
\caption{(Color online). $a)$ The total DOS , $b)$ Ga $s$ (black solid line), $p$ (red dotted line) and  $d$(dashed green line)
projected density of states and $c)$ N $s$ (black solid line), $p$ (red dotted line) and  $d$(dashed green line)
projected density of states for GaN using PAW potentials 
with no Ga-d. The zero of energy corresponds to the valence band maximum.}
\end{figure}

\begin{figure}
\includegraphics[width=5.5in,angle=270]{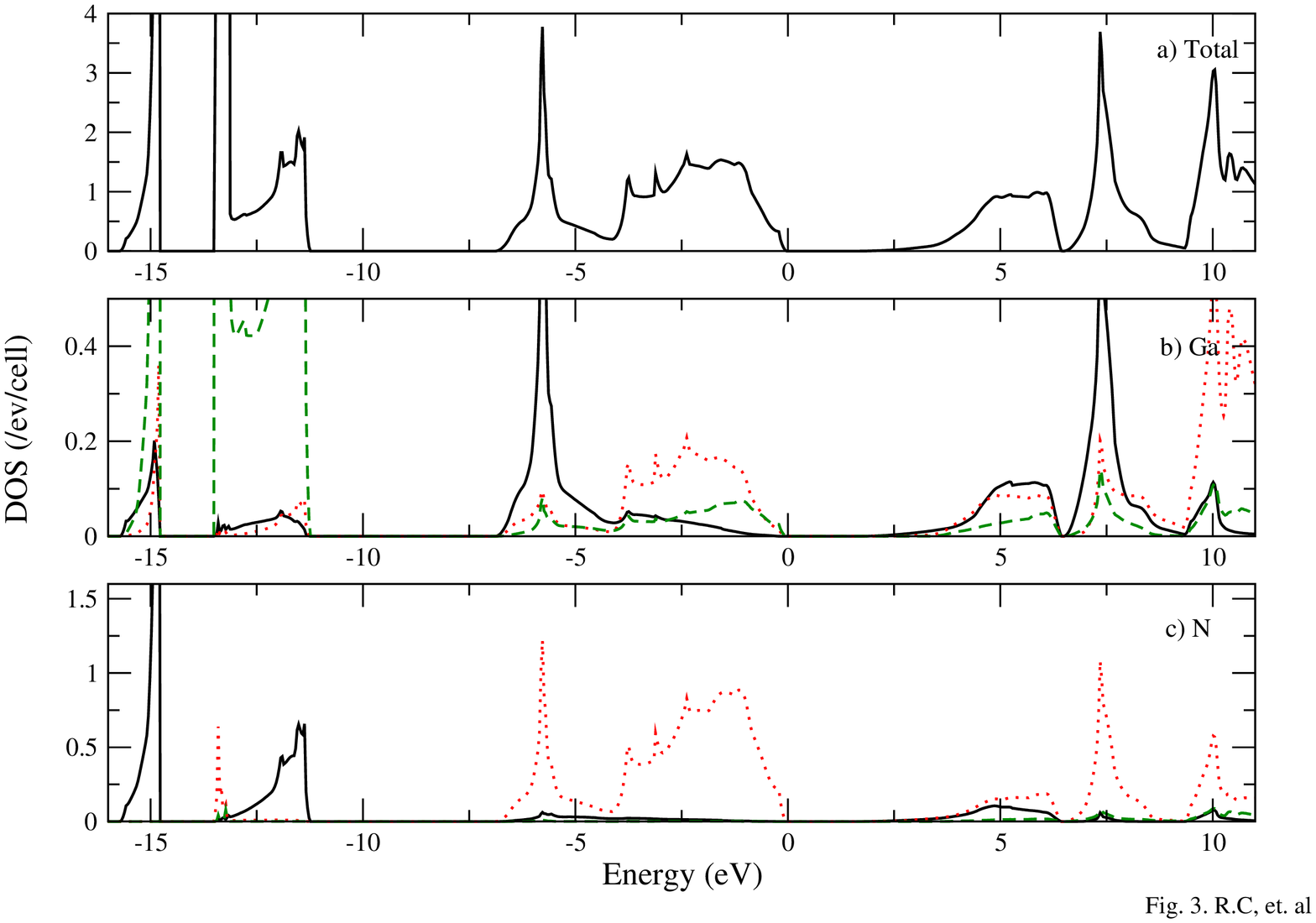}
\caption{(Color online). $a)$ The total DOS , $b)$ Ga $s$ (black solid line), $p$ (red dotted line) and  $d$(dashed green line)
projected density of states and $c)$ N $s$ (black solid line), $p$ (red dotted line) and  $d$(dashed green line)
projected density of states for GaN using PAW potentials 
with Ga-d. The zero of energy corresponds to the valence band maximum. }
\end{figure}

\begin{figure}
\includegraphics[width=5.5in,angle=270]{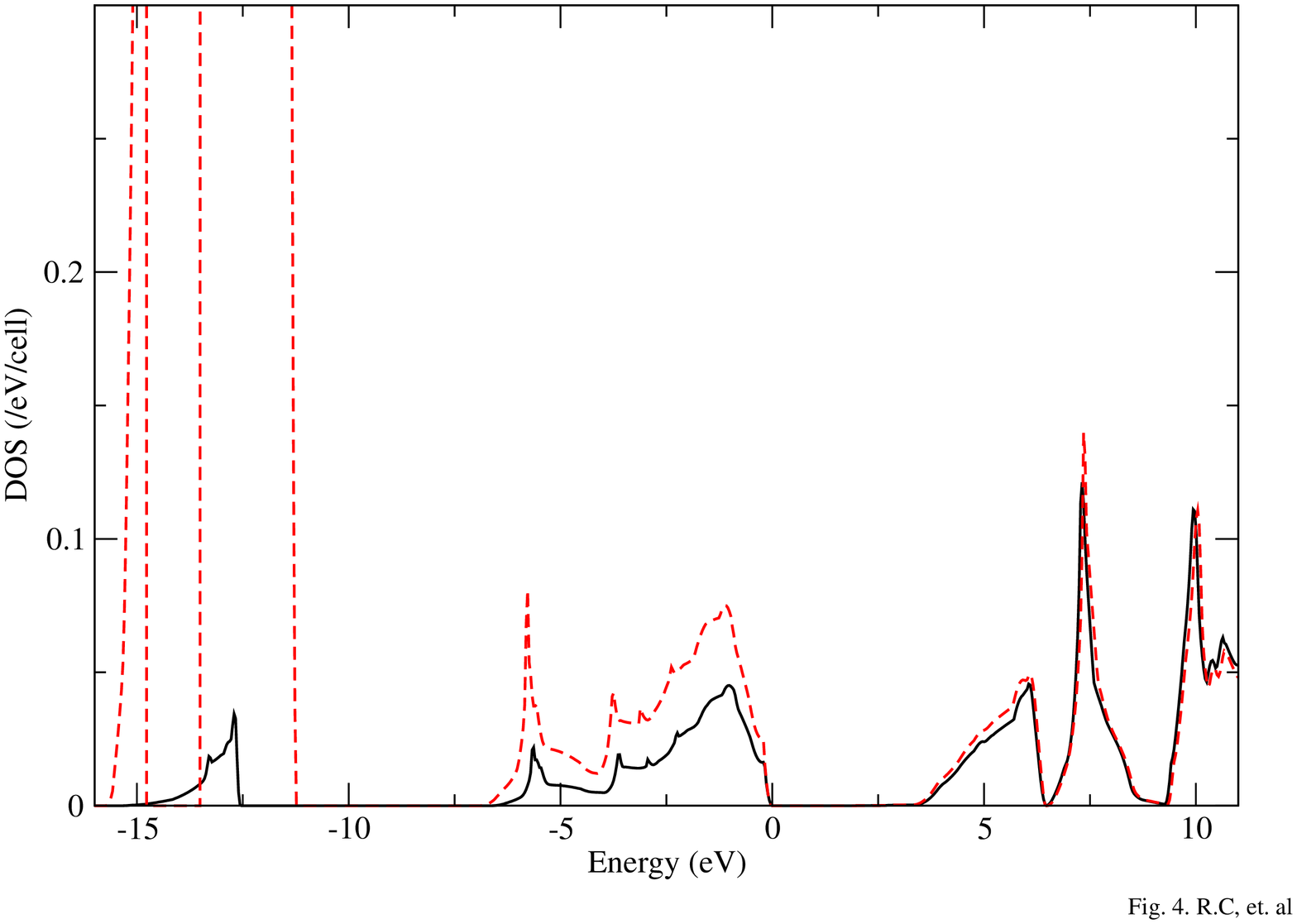}
\caption{(Color online). The Ga $d$ projected density of states for GaN using PAW potentials with (red dotted line) and 
without (black solid line) Ga 3$d$ as a part of the valence band. 
The zero of the energy corresponds to the valence band maximum.}
\end{figure}

\begin{figure}
\includegraphics[width=5.5in,angle=270]{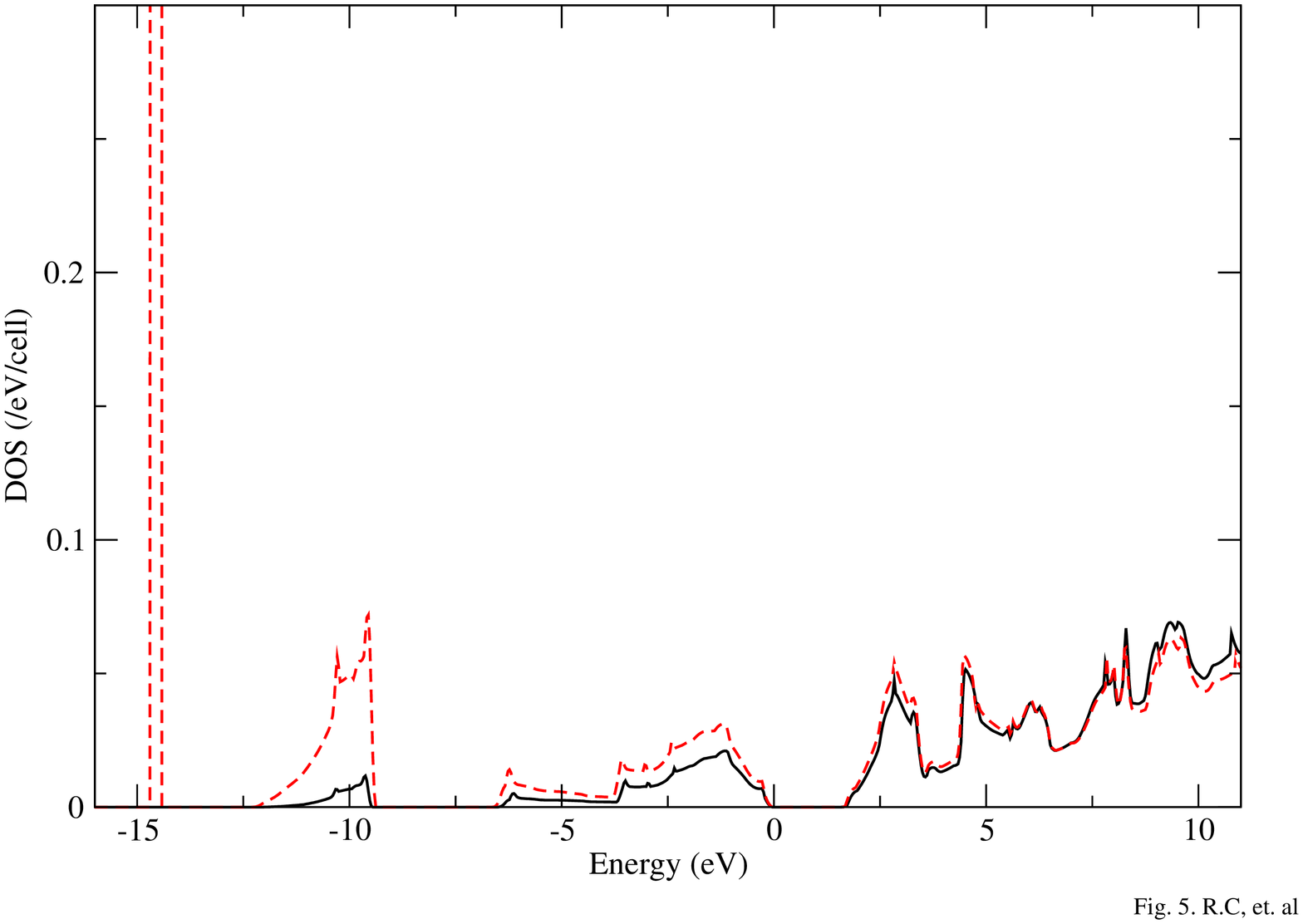}
\caption{ (Color online). The Ga $d$ projected density of states for GaP using PAW potentials with (red dotted line) and
without (black solid line) Ga 3$d$ as a part of the valence band. The 
zero of energy corresponds to the valence band maximum.}
\end{figure}

\begin{figure}
\includegraphics[width=5.5in,angle=270]{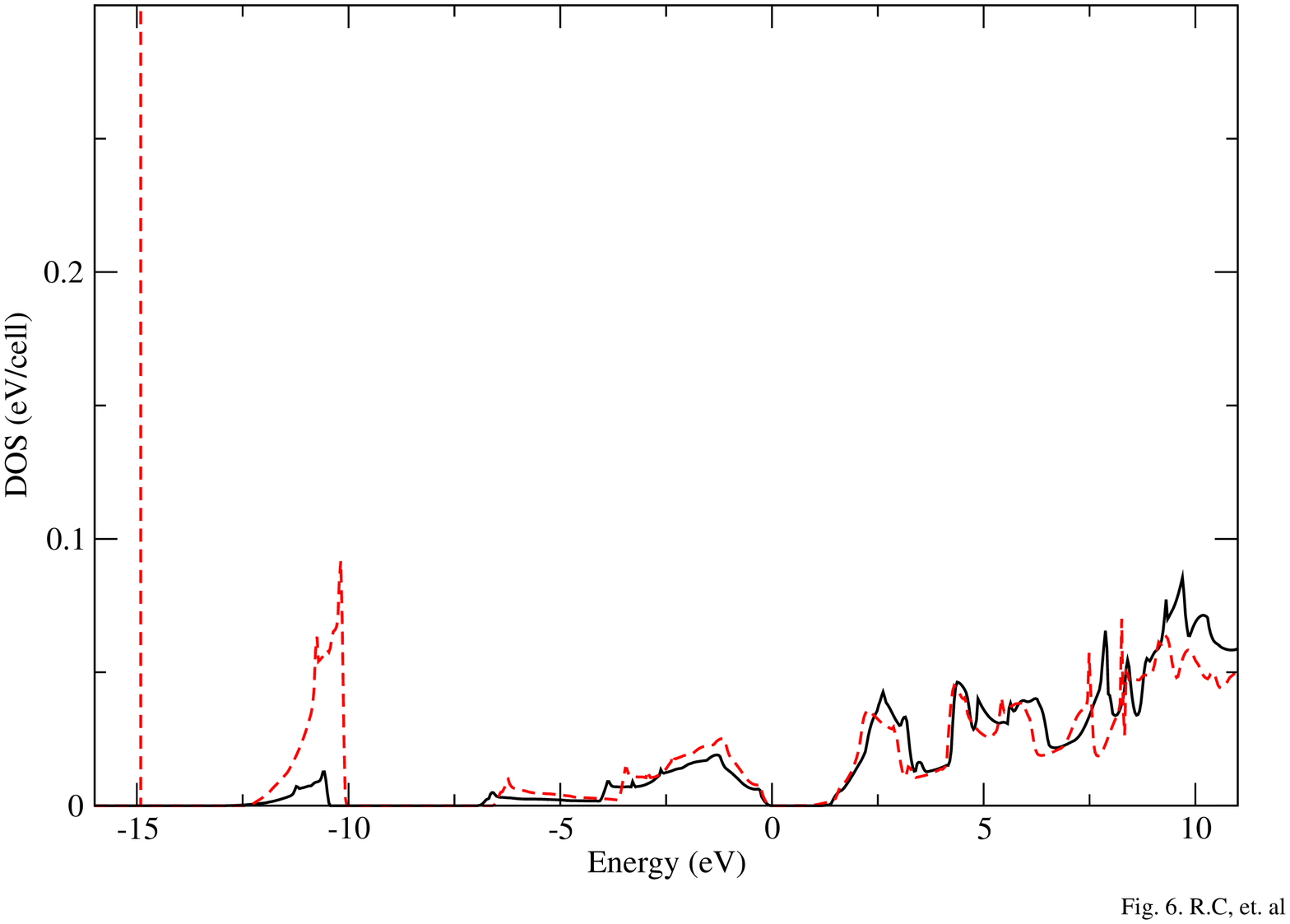}
\caption{ (Color online). The Ga $d$ projected density of states for GaAs using PAW potentials with (red dotted line) and
without (black solid line) Ga 3$d$ as a part of the valence band. The 
zero of energy corresponds to the valence band maximum.}
\end{figure}

\begin{figure}
\includegraphics[width=5.5in,angle=270]{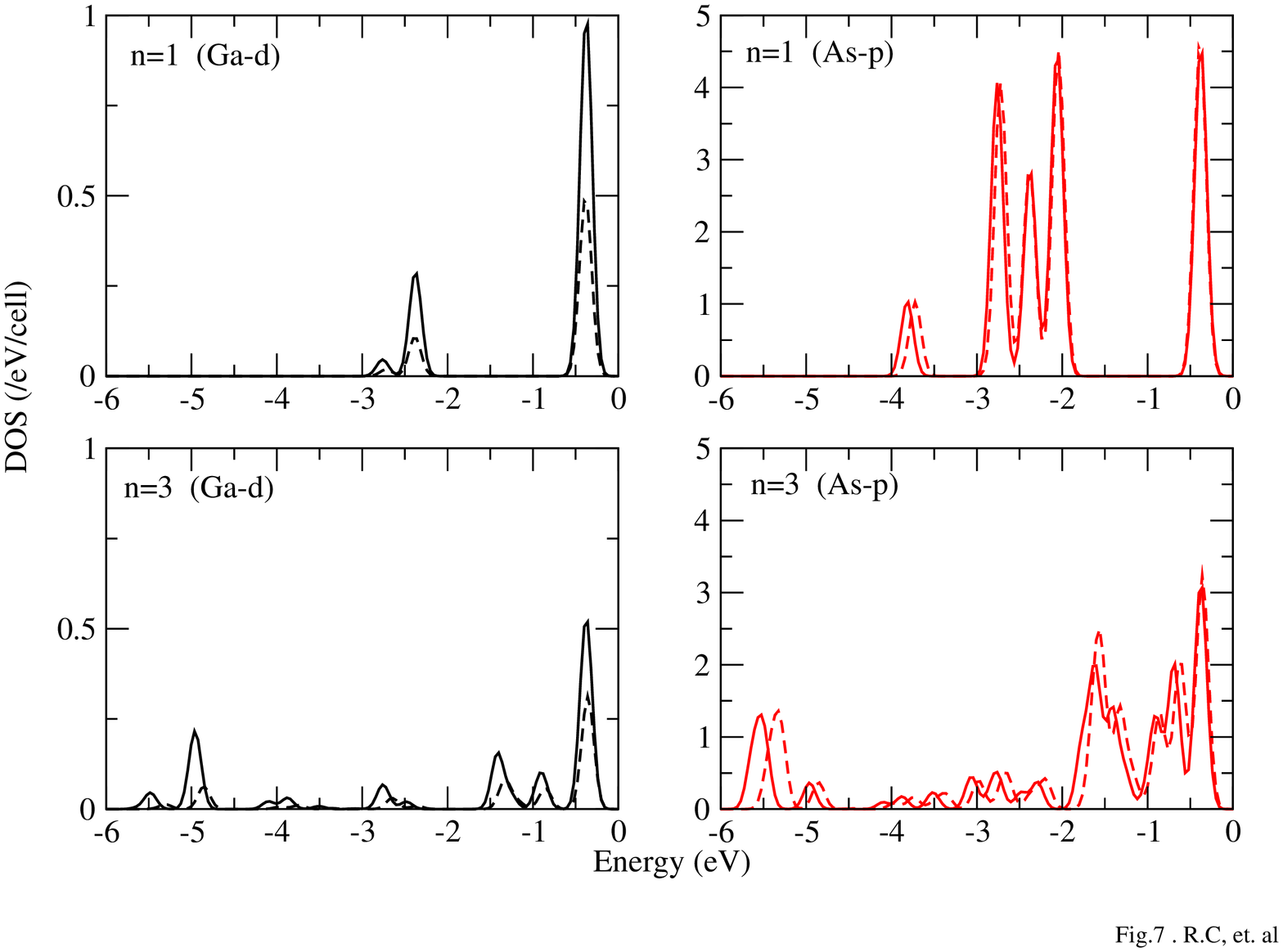}
\caption{(Color online). The Ga $d$ projected density of states (left panel) using PAW potentials with (black solid line)
and without (black dashed line) Ga 3$d$ as a part of the valence band,
the N $p$ projected density of states (right panel) using PAW potentials with (red solid line) and without (red dashed line) Ga 3$d$ 
as a part of the valence band 
for the two cluster sizes n=1 (top panel) and n=3 (bottom panel) 
considered have been shown. The zero of energy corresponds to the 
valence band maximum.}  
\end{figure}

\newpage
\begin{table}
\caption
{The lattice constant, a$_0$ ($\AA$) and  bulk modulus, B (G Pa), variation in GaN with different 
potentials.}
\begin{tabular}{|c|c|c|c|c|c|}
\hline
 & \multicolumn{2}{c} {PAW}  \vline &  \multicolumn{2}{c} {USP} \vline \\
\cline{2-5}
 &  no Ga-d & with Ga-d &  no Ga-d & with Ga-d   \\
\hline
a$_0$ & 4.584 & 4.545 & 4.513 & 4.530 \\
B &   183.63 & 172.59 & 177.33 & 170.03 \\
\hline
\end{tabular}
\end{table}

\begin{table}
\caption
{Calculated structural properties for GaX, X= N, P and As. The lattice constant a$_0$ is in $\AA$, B is the bulk modulus in G Pa.}
\begin{tabular}{|c|c|c|c|c|c|}
\hline
 & \multicolumn{4}{c} {PAW}  \vline \\
\cline{2-5}
 & \multicolumn{2}{c} {no Ga-d} \vline & \multicolumn{2}{c} {with Ga-d} \vline \\
\cline{2-5}
 & a$_0$ & B & a$_0$ & B \\ \hline
GaN   & 4.584 & 183.63 & 4.545 & 172.59 \\
GaP   & 5.532 & 78.74 & 5.504 & 76.70 \\
GaAs & 5.759 & 62.47 & 5.746 & 61.28 \\
\hline
\end{tabular}
\end{table}

\begin{table}
\caption
{Optimised lattice constant in $\AA$ for Ga centered clusters.}
\begin{tabular}{|c|c|c|c|c|c|}
\hline
Cluster size & \multicolumn{2}{c} {PAW}  \vline \\
\cline{2-3}
(n) & no Ga-d & with Ga-d \\  \hline
1 & 4.521 & 4.483 \\
3 & 4.550 & 4.509 \\
\hline
\end{tabular}
\end{table}

\end{document}